\documentclass[journal]{IEEEtran}
\usepackage{multirow}
\usepackage{amssymb}

\usepackage{cite}
\ifCLASSINFOpdf
  \usepackage[pdftex]{graphicx}
\else
\fi

\begin{document}

\title{Hardware implementation of the ring generator with tunable frequency based on electronic neurons}

\author{Nikita~M.~Egorov,
        Marina~V.~Sysoeva, 
        Vladimir~I.~Ponomarenko,
		Ilya~V.~Sysoev
\thanks{The authors are with the laboratory of experimental nonlinear dynamics of Saratov Branch of Kotel'nikov Institute of Radioengineering and Electronics of RAS}
\thanks{Manuscript received \today}}

\markboth{Arxiv.org, \today}%
{Egorov \MakeLowercase{\textit{et al.}}: Ring generator with tunable frequency based on electronic neurons}

\maketitle

\begin{abstract}
Constructing electronic models of neurons has several applications including reproducing dynamics of biological neurons and their networks and neuroprosthetics. In the brain, most neurons themselves are in a non-oscillatory mode, and brain rhythms arise due to their collective dynamics. In this case, very small ensembles of neurons can act as rhythm generators. such ensembles can be constructed and study within the framework of a radiophysical experiment. In this work, a circuit of a ring rhythm generator was created from several (from two to eight) FitzHugh--Nagumo electronic oscillators with electronic synapses (sigmoid coupling function and delay were implemented). Oscillatory modes were shown to be possible in this circuit as a result of collective dynamics, with the frequency being controlled by means of a delay in the synapse and/or changing the number of elements in the ring. For some parameter values, there were multistability modes, when the implementation of a specific oscillatory regime was determined by initial conditions or could be achieved by a short-term external driving. The constructed generator well models the switching of the main frequency in brain local field potentials at limbic epilepsy, but can be used independently as well.
\end{abstract}


\IEEEpeerreviewmaketitle

\section{Introduction}
The construction of models reflecting the functioning of real neurons and their groups is of both scientific and practical technical interest. In particular, the concept of a central pattern (rhythm) generator is actively developing in robotics \cite{Lee_etal_Neurocomputing2007,Lodi_IEEETrans2020,Kurkin_Chaos2022}. This rhythm is necessary for the realization of simple movements which are characteristic of living organisms. When modeling pathological modes of brain functioning, in particular, when modeling epilepsy, the question of the basic rhythm generation is also of a large importance. We have attempted to describe the occurrence and evolution of the main oscillation frequency in the hippocampus during limbic epilepsy (see the modern epilepsy classification in \cite{Scheffer_etal2017}) using a ring of a small number of radiophysical oscillators constructed from physiological reasons (mostly studies of rat models were used \cite{Cavalheiro_1995,Coulter_etal_BrainPath2002}). Compared with mathematical modeling, this technique allows us to approach a biological experiment by a number of criteria: from the point of view of the specifics of measurements, from the point of view of the non-stationarity (thermal heating) of the circuit parameters and their non-identity. At the same time, electronic circuits using microcontrollers such as Arduino \cite{Tagne_etal2022} are another (next to SPICE simulators) intermediate step between fully analog modeling and numerical solution of equations.

When constructing electronic models of neurons, the main approach is the schematic reproduction of mathematical models \cite{MahowaldDouglas_Nature1991,RascheDouglas_SignalProc2000,vanSchaik_NeuralNetworks2001}. From the many variants of mathematical models of biological neurons \cite{Dmitrichev_AND2018}, the FitzHugh--Nagumo neuron model was chosen \cite{FitzHugh1961,Nagumo1962}, which is a dimensionless simplified version of the Hodgkin--Huxley model \cite{HodgkinHuxley1952}, reproducing the basic properties of excitation waves. The main reason for researchers' interest in this model is the simplicity of implementing nonlinear functions, making possible to assemble an electronic model on the simplest elements \cite{Linares_etal_IEEE1991,Binczak_etal_ElectronLett2003,ZhaoKim_IEEEProc2007} or to implement an ensemble of ten or more elements \cite{Egorov_ND2022} relatively quickly .

Earlier in the paper \cite{Kulminskiy_ND2019}, a hardware implementation of a simplified FitzHugh--Nagumo neuron with one bifurcation parameter was implemented. The SPICE--model consisting of 14 simplified FitzHugh--Nagumo neurons connected linearly was built in \cite{Egorov_TPhL2021}. The scaling of dynamics with increase of number of neurons in the model was demonstrated in \cite{Egorov_AND2021}. Variability inherent for biological systems due to different in detail coupling architecture was also shown. Then, eight hardware circuits consisting of 14 simplified FitzHugh--Nagumo neurons were implemented \cite{Egorov_ND2022}. The electronic experiment showed that the implemented circuits are able to demonstrate the desired behavior --- long-term quasi-regular transients reproducing various characteristics of epileptiform activity. The similar transient dynamics had been previously found in mathematical models \cite{Kapustnikov_MatBio2020,Kapustnikov_CNSNS2022}. Further, a circuit of a complete FitzHugh--Nagumo neuron with two bifurcation parameters $a$ and $b$ and a circuit of a chemical synapse mathematically representing a sigmoid function were developed in \cite{Egorov_SPIE2022}. In \cite{Egorov_CSF2022} it was shown that in two hardware complete FitzHugh--Nagumo neurons connected by sigmoid couplings, different scenarios of oscillation occurrence were possible, including saddle node cycle bifurcation, leading to the appearance of highly nonlinear limit cycles of large amplitude. Long-living transients have been detected near these bifurcations.

Here, we base our research on the individual neuron model developed in \cite{Egorov_CSF2022}. However, it does not completely satisfy our demands since the delay naturally occurs in the synapse when a signal is transmitted between the axon and the dendrite as a result of finite speed of the ion transport. The delay can have a significant impact on the network dynamics \cite{Wang_PhysA2010}. Therefore, the previous model was improved by adding a delay in the coupling \cite{Egorov_AND2023}:
\begin{eqnarray}
\varepsilon \dot u_i (t) &=& u_i(t)- c_i u_i^3(t)-v_i (t) + \sum_{j \neq i}^{} k_{ij} h \left(u_j (t - \tau) \right), \nonumber\\
\dot v_i (t) &=& u_i (t) + a_i - b_i v_i (t), \label{eq:FHN}\\
h(u) &=& \frac{1+\tanh(u)}{2}, \nonumber
\end{eqnarray}
where $u$ is a dimensionless variable analogous to the transmembrane potential in biological excitable tissue; $v$ is a dimensionless variable similar to slow recovery current; $t$ is dimensionless time; $\varepsilon$ is an inertia parameter; $a$ and $b$ are dimensionless parameters that control the dynamics of individual neuron; $c$ is an integration constant (in our papers always $c=1/3$); $k$ is a coupling coefficient, while the coupling is implemented in the form of an hyperbolic tangent with shift $h$; $\tau$ is a delay time.

Four variants of tunable analog delay were compared in \cite{Egorov_AND2023}. The standard ``DELAY'' component from the National Instruments Multisim SPICE simulator was used as a reference. We also considered a first-order all-pass filter with a potentiometer, a Bessel filter with a potentiometer, and sequentially connected Bessel filters, each of those providing a fixed delay time. As a result, it was decided to focus on one Bessel filter with a potentiometer as a preferable implementation. The SPICE--circuit of a ring generator consisting of 20 neurons with improved synapses was created in \cite{Egorov_AND2023}. The key feature of this generator was that its frequency could be stated in three ways: by changing the delay time (smooth tuning is available in a wide range), by changing the number of elements in the network (frequency tuning is carried out stepwise), by varying the frequency of an external driving (under conditions of multistability, coexisting regimes with multiple frequencies can be realized).

The purpose of this work is to construct the hardware implementation of a ring generator, the theoretical study of which was performed in \cite{Egorov_AND2023}.

\section{Analog generator}

\begin{figure}[h!]
	\centering
	\includegraphics[width=\linewidth]{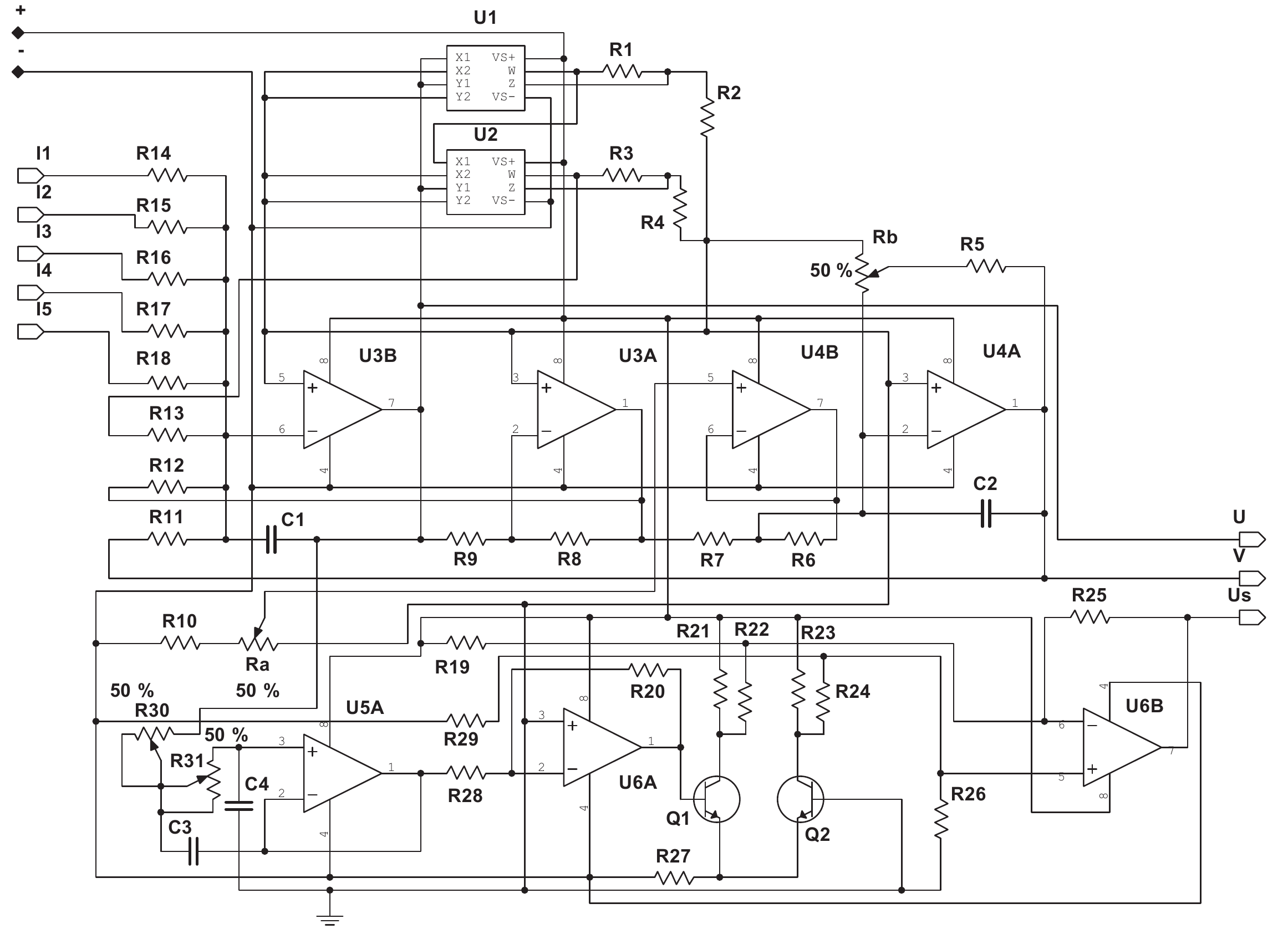} \\
	\caption{Circuit diagram of a single complete FitzHugh--Nagumo neuron with synapse.\\
	Neuron: $R1 = R3 = 1~\mathrm{k\Omega}$, $R2 = 9~\mathrm{k\Omega}$, $R4 = 2.333~\mathrm{k\Omega}$, $R5 = 51~\mathrm{k\Omega}$, potentiometer $R_b = 4.7~\mathrm{M\Omega}$, $R6 = R7 = R8 = R9 = R11 = R12 = R13 = 100~\mathrm{k\Omega}$, $R10 = 5~\mathrm{k\Omega}$,  potentiometer $R_a = 1~\mathrm{k\Omega}$, $R14 - R18$ depends on coupling strength $k$, $C1 = 1$~nF, $C2 = 0.01$~uF, U1, U2 are multipliers of the type AD633, and U3, U4 are amplifiers of the type AD822.\\
	Delay in the form of a Bessel filter with potentiometers: $R30 = R31 = 50~\mathrm{k\Omega}$, $C3 = 5.6$~nF, $C4 = 3.9$~nF, U5A is an amplifier of the type LM358AD.\\
	Sigmoid function: $R19 = R29 = 300~\mathrm{k\Omega}$, $R20 = 0.51~\mathrm{k\Omega}$, $R21 = R23 = 1~\mathrm{k\Omega}$, $R22 = R24 = R28 = 10~\mathrm{k\Omega}$, $R25 = R26 = 5.1~\mathrm{k\Omega}$, $R27 = 2~\mathrm{k\Omega}$, Q1, Q2 are bipolar junction transistors of the type 2N1711, U6 is an amplifier of the type NE5532AI.
	\label{fig:Neuron}}
\end{figure} 

A circuit diagram of the complete FitzHugh--Nagumo electronic oscillator with synapse is shown in Fig.~\ref{fig:Neuron}. In contrast to the mathematical model, see Eq.~(\ref{eq:FHN}), the parameters of the radio engineering circuit have dimensions. The time-scale parameters have the values $E = R11C1$ and $T = R7C2$. Let the dimensional values of the mathematical dimensionless variables $u$ and $v$ be denoted in the scheme as $U$ and $V$ correspondingly. The parameter $\varepsilon$ from Eq.~(\ref{eq:FHN}) is calculated as $\varepsilon = E/T$. The parameters $c = (R3+R4)/R3$ and $b = R6/(R5+R_b \cdot \frac{B}{100\%})$ (for $B$ percentage of the potentiometer $R_b$ was used) are scaling factors at $U$ and $V$ respectively. Coupling coefficient $k$ is calculated as $k = R13/R_{IN}$, where $R_{IN}$ is the nominal value for one of the input resistors from $R14$ to $R18$. The parameter $a$ is set by the voltage at the ``+'' clamp of the amplifier U3B. The total voltage drop on a series-connected resistor $R10=5$~k$\mathrm\Omega$ together with a potentiometer $R_a=1$~k$\mathrm\Omega$ is $U_a = 15$ V, i.\,e.\ the voltage drop on the entire potentiometer is 2.5~V. In particular, if the potentiometer is set to $A=0\%$, the voltage 2.5~V is set to ``+'' input of U3B, and if the potentiometer is set to $A=100\%$, this voltage is zero. So, the parameter $a$ can be calculated using $A$ measured in percents of potentiometer resistance as follows: $a = 2.5 \cdot \left(1-\frac{A}{100\%}\right)$.

The circuit contains two analog multipliers U1 and U2 and two dual operational amplifiers U3 and U4. Elements U4B and U3A are integrators. They allow to obtain $U$ and $V$, respectively. Element U4A is an inverter. It allows to obtain $-U$. Element U3B is a repeater. The multipliers U1 and U2 perform the cubic transformation according to the formula (\ref{eq:FHN}).

The chemical synapse circuit consists of two parts: a circuit simulating an analog delay, and a circuit implementing a sigmoid function (radio engineering implementation of a hyperbolic tangent). The first circuit contains a second-order Bessel filter based on the operational amplifier U5A. The second circuit contains a dual operational amplifier U6 and two bipolar transistors Q1 and Q2. The inverting amplifier U6A has a gain of $0.05$, the differential amplifier U6B has a gain of $0.5$.

\begin{figure*}
	\centering
	\includegraphics[width=\textwidth]{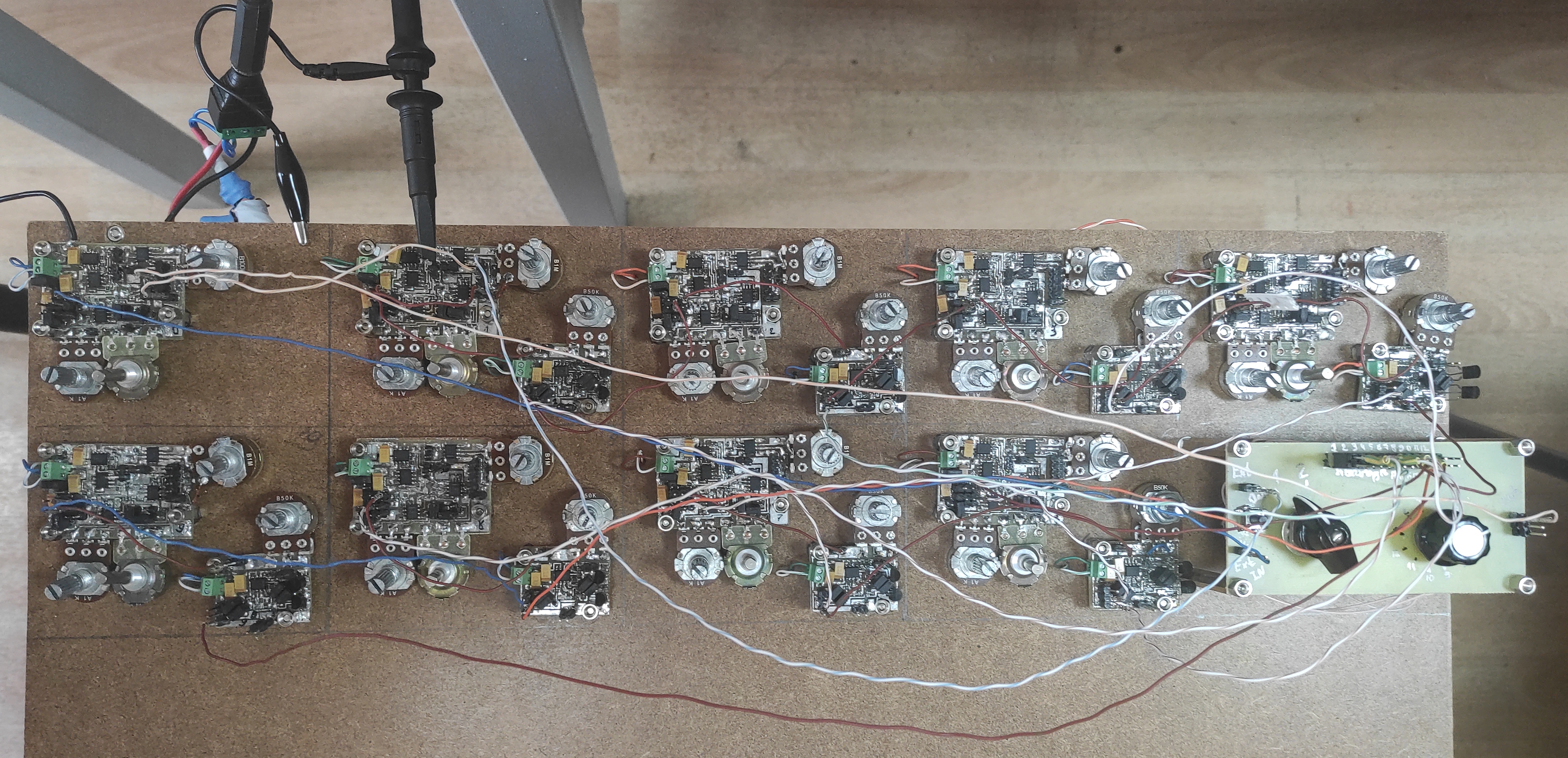} \\
	\caption{Photograph of the hardware implementation of an electronic ring generator of neuron-like activity with tunable frequency. N0 --- external excitatory input ($a_{\rm input} = 0.9$, $b_{\rm input} = 0.09$, $k_{\rm input} = 1.0$); N1--N8 --- the neurons of the ring ($a_{\rm ring} = 1.1$, $b_{\rm ring} = 0.09$, $k_{\rm ring} = 1.0$).
	\label{fig:Circuit}}
\end{figure*} 

The physical implementation of the ring generator consists of eight basic neurons N1--N8 and one additional neuron of the external input N0 (Fig. ~\ref{fig:Circuit}). The basic neurons are in subthreshold mode with parameters ${a_{\rm ring}} = 1.1$ and ${b_{\rm ring}} = 0.09$. The external input is in an oscillatory mode with parameters ${a_{\rm input}} = 0.875$ and ${b_{\rm input}} = 0.08$. Driving from it is provided, if necessary, by pressing the button (bottom right in the Fig.~\ref{fig:Circuit}). The coupling coefficient inside the ring and the coupling coefficient of external (starting) neuron were taken the same ${k_{\rm ring}} = {k_{\rm input}} = 1.0$. The number of neurons in the ring was changed by means of the rotary switch from one to eight discretely.

When neurons are closed in a ring, a coupling occurs between them with a delay time, which corresponds to a delay in a chemical synapse in a real biological neuron caused by a finite times required for ion transport through the synapse. As a result, an oscillatory attractor appears in the circuit, while a separate neuron has a single stable fixed point at the considered parameters. However, this oscillatory attractor (usually cycle) often coexists with a stable equilibrium, since it arises rigidly, not due to the Andronov--Hopf bifurcation, but as a result of the saddle-node bifurcation of the cycle \cite{YuriAKuznetsov_book,Gonchenko_etal_AND2019} (in the two-dimensional case --- this is the birth of a cycle from the condensation of phase trajectories \cite{RabinovichTrubetskov_book1989}). This mechanism for mathematical models of FitzHugh--Nagumo neuron networks was studied in \cite{Kapustnikov_CNSNS2022}. It is responsible both for the formation of the attractor and for long-term transient dynamics occurring near the bifurcation point. Approaching the cycle depends on the initial perturbation, for example, when an electronic key is closed, or can be carried out by a short-term external driving. In this case, the model can go through a bifurcation of the birth (or death) of a cycle directly during the experiment due to changes in the parameters of electronic neurons due to heating.

\section{Results}
Data were recorded with a two-channel oscilloscope (bandwidth up to 100 MHz, sampling frequency 500 MHz per channel, quantization bit length 8 bits) from neurons N1 and N5. The hardware implementation of the ring generator generated oscillations without additional external input in contrast to the SPICE--circuit \cite{Egorov_AND2023}. Most likely, switching the positions of the rotary switch itself can be considered a short-term external driving. However, during the experiment, with each set of parameters, the behavior was recorded both without external influence and after a short-term signal from neuron N0 was given.

\begin{figure}[h!]
\centering
	\begin{minipage}[h]{0.49\linewidth}
		{\center{\includegraphics[width=0.98\linewidth]{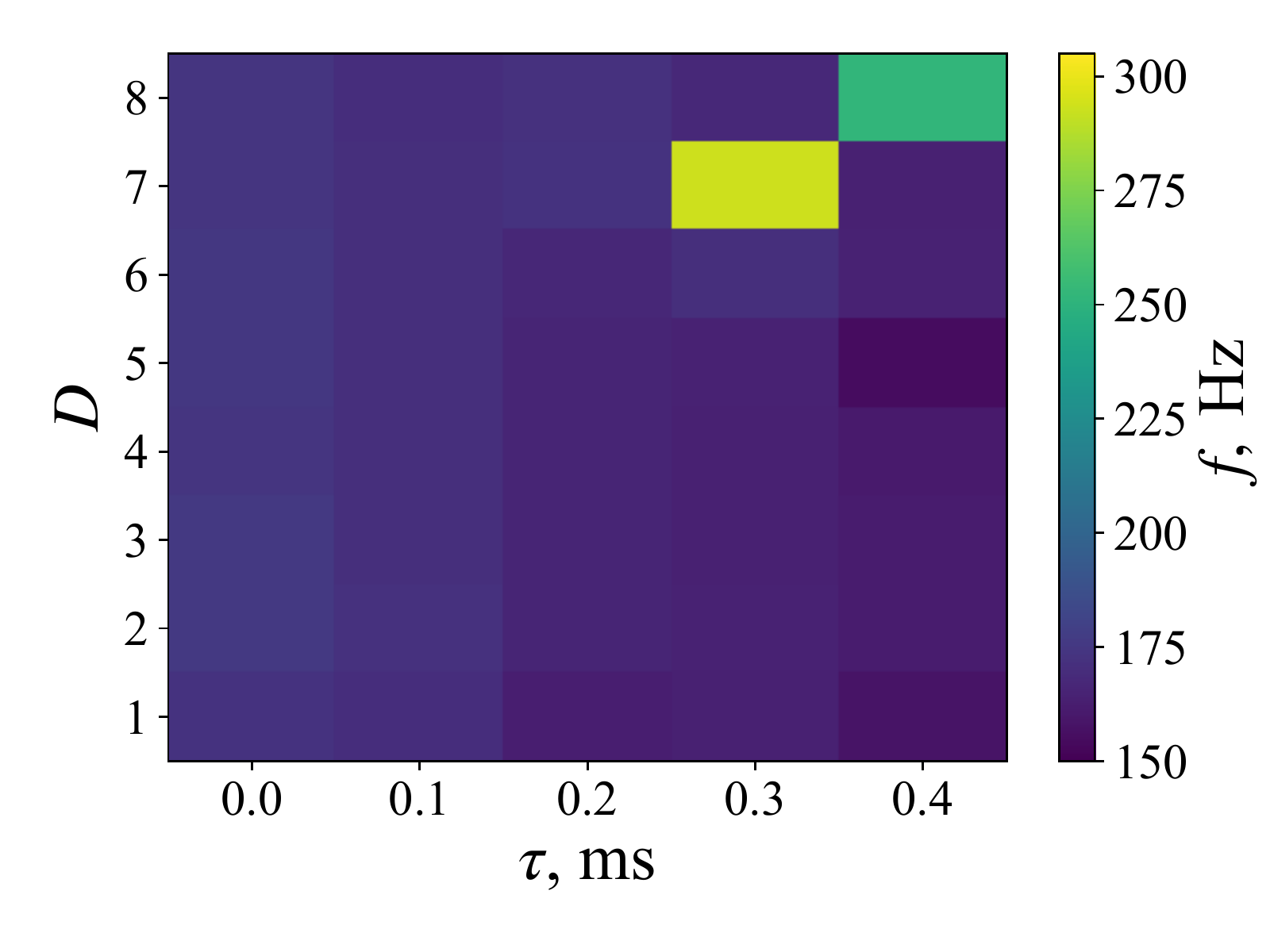}}}
	\end{minipage}
	\hfill
	\begin{minipage}[h]{0.49\linewidth}
		{\center{\includegraphics[width=0.98\linewidth]{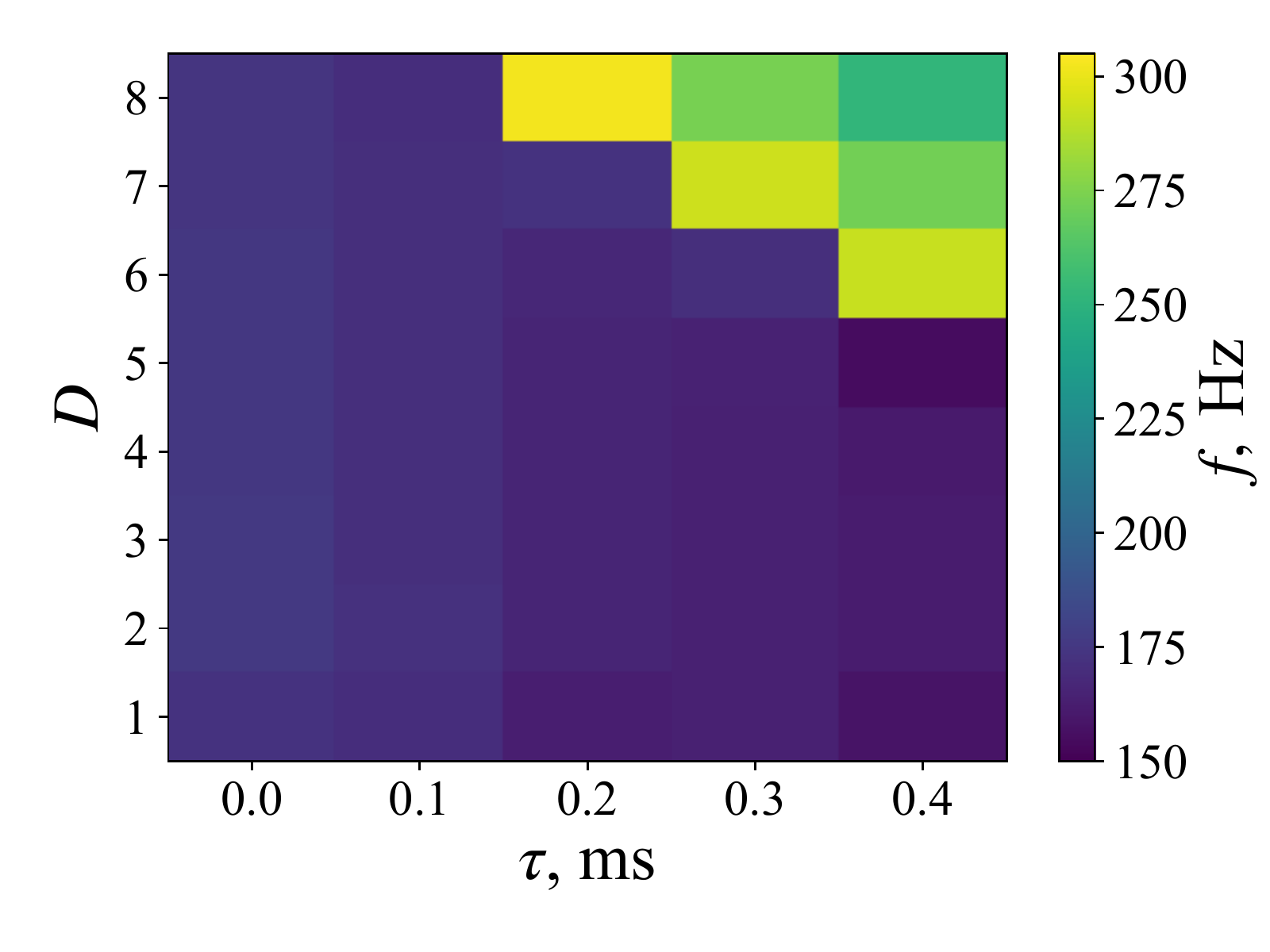}}}
	\end{minipage}
	\vfill
\hspace{-0.05\linewidth}(a)\hspace{0.5\linewidth}(b)
\caption{Dependencies of the main oscillation frequency $f$ in the circuit on the number of elements in the ring $D$ and on the delay time $\tau$: (a) without the driving; (b) after the external driving. The color indicates the oscillation frequency occurring in the ring.
\label{fig:Diagram}} 
\end{figure}

Dependencies of the main oscillation frequency $f$ in the circuit on the number of elements in the ring $D$ and on the delay time $\tau$ is shown in Fig.~\ref{fig:Diagram}. The diagram obtained without external driving is shown in Fig.~\ref{fig:Diagram} (a). The diagram after the external driving is shown in Fig.~\ref{fig:Diagram} (b). These two diagrams are mostly similar, except for points in the right upper corner. This indicates bistability (actually, multistability, since the steady state can be also available for some specific initial conditions). To calculate the main oscillation frequency from time series, the oscillation period $T_1$ of neuron N1 was estimated and the frequency was estimated as $1/T_1$. It is worth emphasizing that the abscissa shows the delay $\tau$, which occurs in a separate circuit simulating an analog delay line, and not the total delay in the circuit, which is summed from all the delays in the circuit. The delay (phase shift) that occurs in the neuron circuit due to the inertia in each individual neuron was not taken into account when plotting the figure.

\begin{figure}[b!]
	\begin{minipage}[h]{0.49\linewidth}
		{\center{\includegraphics[width=0.98\linewidth]{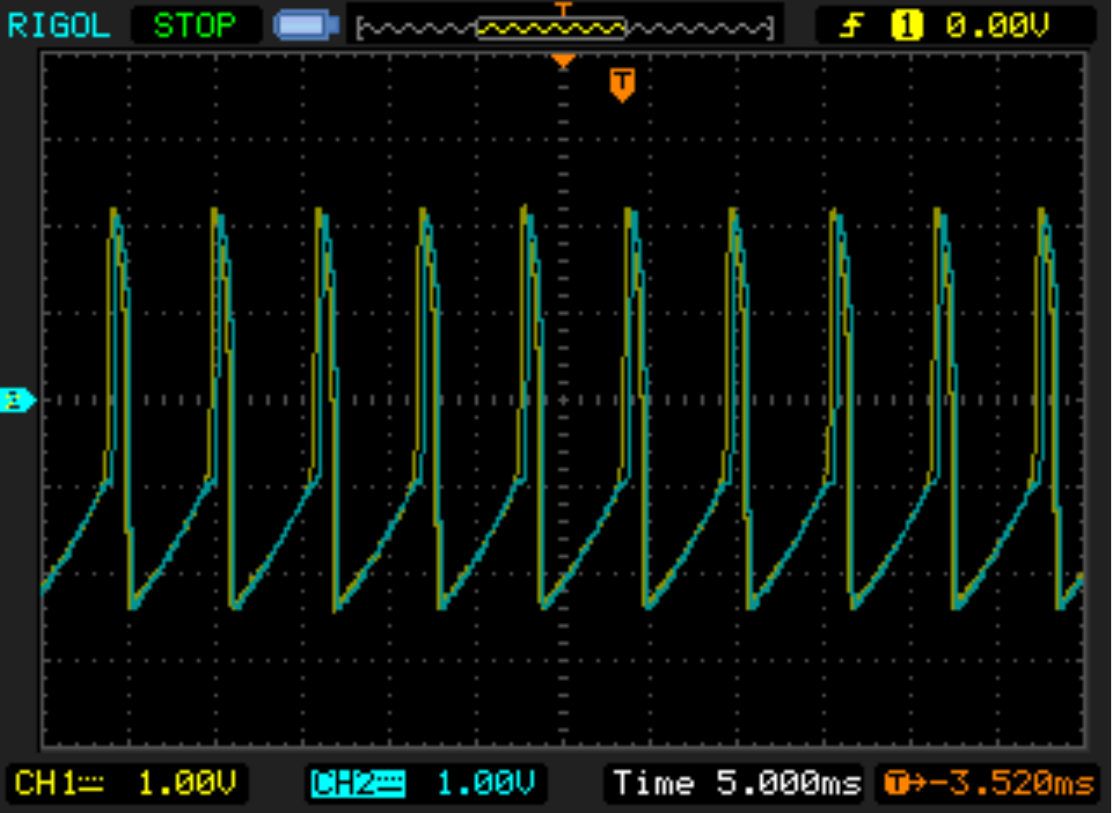}}}
	\end{minipage}
	\hfill
	\begin{minipage}[h]{0.49\linewidth}
		{\center{\includegraphics[width=0.98\linewidth]{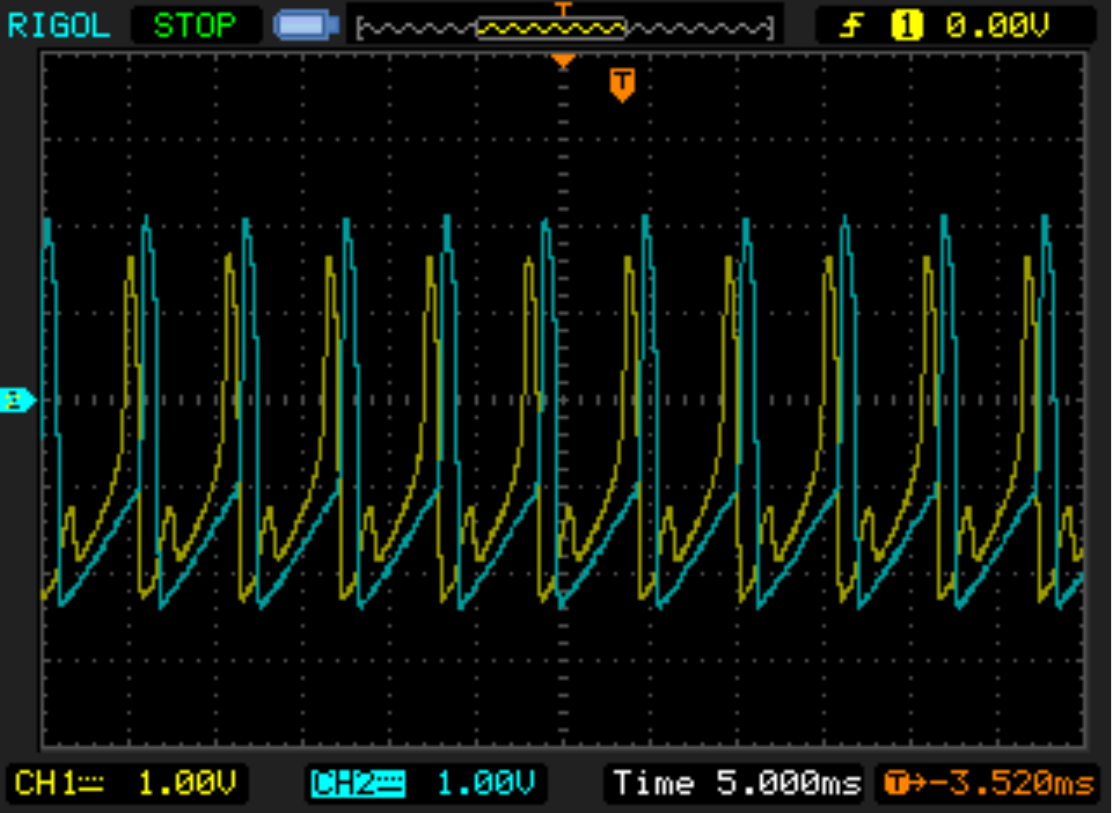}}}
	\end{minipage}
	\vfill
	\hspace{0.2\linewidth}(a)\hspace{0.5\linewidth}(b)
	\vfill	
	\begin{minipage}[h]{0.49\linewidth}
		{\center{\includegraphics[width=0.98\linewidth]{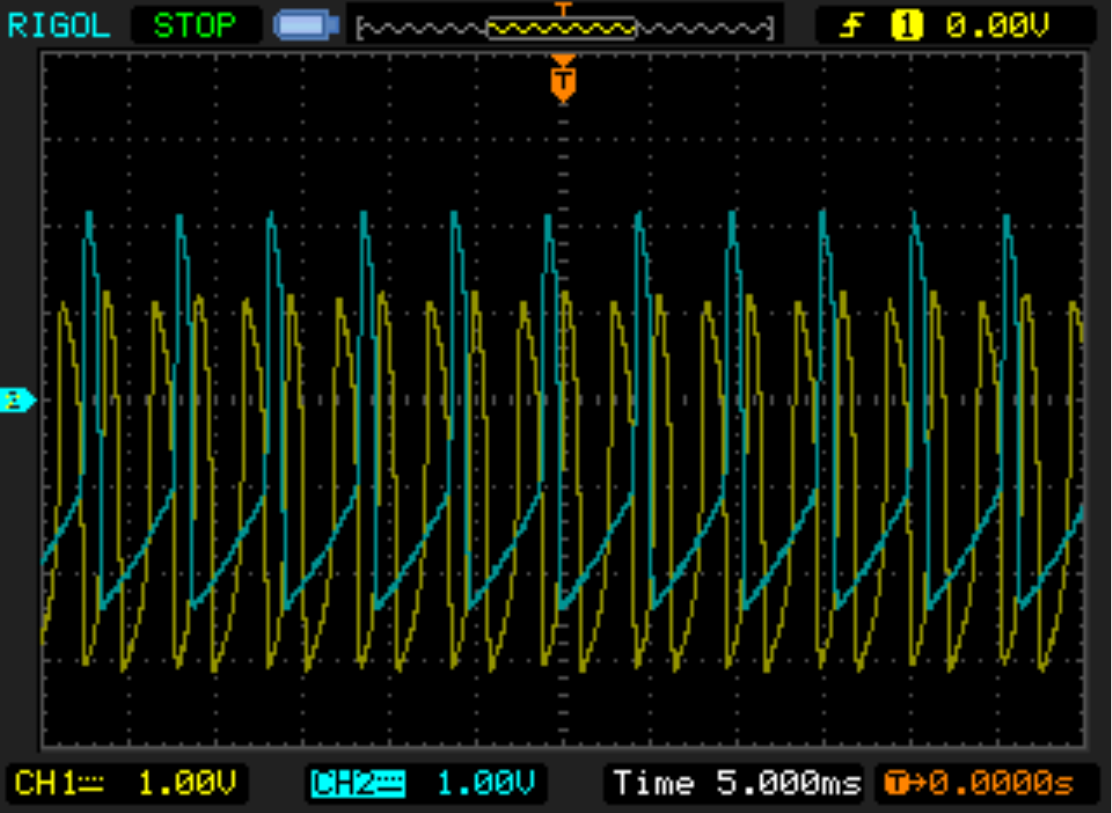}}}
	\end{minipage}
	\hfill
	\begin{minipage}[h]{0.49\linewidth}
		{\center{\includegraphics[width=0.98\linewidth]{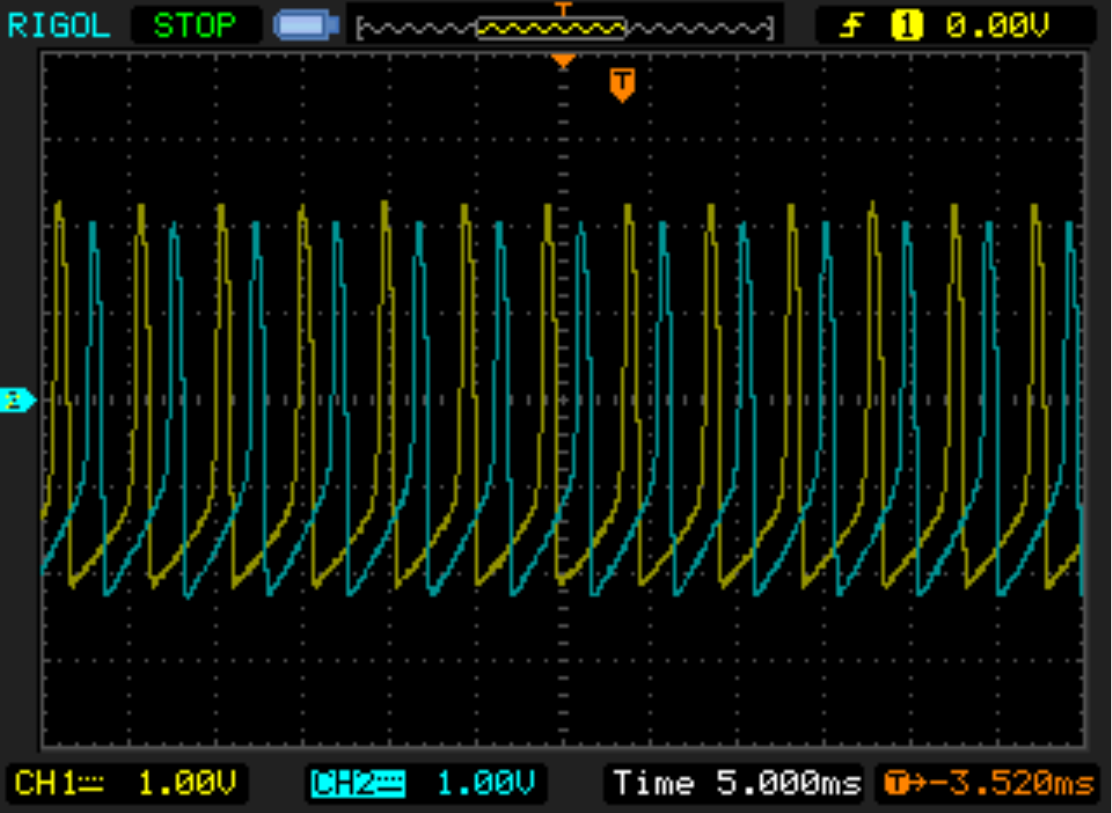}}}
	\end{minipage}
	\vfill
	\hspace{0.2\linewidth}(c)\hspace{0.5\linewidth}(d)
	\caption{Time series: (\textit{a}) inphase oscillations $\tau=0.0$, $D=7$; (\textit{b}) multimodal oscillations $\tau=0.2$, $D=7$; (\textit{c}) multimodal oscillations $\tau=0.3$, $D=7$; (\textit{d}) lag-synchronous oscillations $\tau=0.4$, $D=7$. Yellow line corresponds to N1. Blue line corresponds to N5. \protect\\ \label{fig:Multi}}
\end{figure}

Three types of behavior were found in the circuit:
\begin{itemize}
	\item inphase oscillations (fig.~\ref{fig:Multi} (a));
	\item multimodal oscillations (fig.~\ref{fig:Multi} (b) and (c));
	\item lag-synchronous oscillations (fig.~\ref{fig:Multi} (d)).
\end{itemize}

Inphase oscillations were observed at small values of the delay time $\tau=0.0$ and $\tau=0.1$ for any investigated number of neurons in the ring, as well as for $\tau=0.2$ for $1 \leqslant D \leqslant 6$, $\tau=0.3$ at $1 \leqslant D \leqslant 5$. At this kind of oscillation activity the ring is a single synchronously oscillating system. In fact, the system dimension is effectively reduced to the dimension of one neuron oscillating with the frequency indicated in Fig.~\ref{fig:Diagram}. The inphase oscillation frequency does not depend on the number of neurons $D$ at a fixed $\tau$. But if we fix $D$, then we can see a clear drop in the main oscillation frequency with an increase in the delay time $\tau$.

Multimodal oscillations were observed for $\tau=0.2$ and $D=7$, $\tau=0.3$ and $D=6$, $\tau=0.4$ and $1 \leqslant D \leqslant 5$. In fact, this is a class of modes with smooth transition between them rather than a single regime. When multimodal oscillations appear initially, they look like in fig.~\ref{fig:Multi} (b). If we consider neurons as separate non-autonomous dynamical systems, some of them undergo period halving bifurcation \cite{Cascais_PLA1983} (see neuron N1 oscillations), while others remain in period 1 mode (see neuron N5 oscillations). As a rule, neurons with different oscillation modes alternate in the ring. Gradually, with an increase in $D$, the oscillatory mode changes to something like presented on the fig.~\ref{fig:Multi} (c). The half-period mode becomes more homogeneous, but no additional bifurcation occurs.

In the present experiment, lag-synchronous oscillations \cite{Rosenblum_PRL1997} were observed without an external driving only in two cases: for $\tau=0.3$ and $D=7$ and for $\tau=0.4$ and $D=8$. After the external driving these oscillations were observed for six combinations of $(\tau, D)$: (0.2, 8), (0.3, 7), (0.3, 8), (0.4, 6), (0.4, 7), (0.4, 8). This mode can be implemented only with sufficiently large delays and large number of oscillators (upper right corner in Fig.~\ref{fig:Diagram}). This mode, along with the non-oscillation mode, was the main one in the mathematical model \cite{Sysoev_NODYCON2022} and SPICE--model \cite{Egorov_AND2023}. At the same time, this mode in the hardware circuit can be observed both by increasing the number of elements and by increasing the delay between elements. Probably, if it were possible to physically implement more neurons, then lag-synchronous oscillations would appear at $\tau=0.2$. Similarly, lag-synchronous oscillations could be observed at a smaller $D$, for example, at $D = 5$, if we increase $\tau$. The main frequency of lag-synchronous oscillations behaves as predicted in the mathematical model and shown in the SPICE--model. With an increase in $\tau$ and $D$, the frequency $f$ decreases, which is especially clearly seen in Fig.~\ref{fig:Diagram} (b).

\section{Conclusion and discussion}
The novelty of this study is as follows. First, the electronic model of an individual neuron was taken from \cite{Egorov_CSF2022}, but a delay was introduced in the coupling, which simulates a real delay in a chemical synapse. The delay was implemented in the form of a tunable Bessel filter, where one of the resistors was replaced by a potentiometer. The delay in the synapse significantly expands the set of modes which can exist in a network of connected electronic neurons. Second, while developing the scheme, we tried to reproduce the results of previous work on mathematical \cite{Sysoev_NODYCON2022} and SPICE \cite{Egorov_AND2023} modeling. We aimed to obtain a specific mode of behavior in a unidirectionally connected ring of neurons, in which the frequency of periodic pulse oscillations is controlled by a delay in couplings and depends on the number of elements. This was achieved for sufficiently large delay values for six or more neurons (in total, in six different combinations of the number of elements $D$ and the delay time $\tau$). Third, in addition to the initial theoretically predicted switching of the main oscillation frequency with a change in the delay time and the number of neurons in the network, the realized circuit also demonstrated switching between different modes (inphase, multimodal and lag-synchronous oscillations), which was not observed in the mathematical and SPICE models.

The hardware implementation of neurons allows to partially reproduce the difficulties inherent in the study of biological neurons. From our point of view,  this is the main advantage over the mathematical modeling. All neurons and synapses are not identical in parameters, they do not perfectly repeat the mathematical model, their parameters change over time, including due to heating. Therefore, the fact of detection of the desired modes indicates their great structural stability. This means that in a biological system, where the role of non-identity and non-stationarity is even greater, one can still expect to observe such modes. The role of the studied regimes can be very great for the formation of the limbic epilepsy basic rhythm. According to modern concepts, a very small network of hippocampal neurons can be responsible for it \cite{PazHuguenard_NatNeurosci2015}. In contrary to microscopic silicon implementations of neurons \cite{Linares_etal_IEEE1991,Folowosele_etal_IEEE2011} which are preferred if the large ensemble is desired, the parameters of the macroscopic models \cite{Binczak_NN2006,Kulminskiy_ND2019,Egorov_CSF2022} can be well controlled. This makes possible specifying parameters for a particular neuron type which is significant for microcircuit modeling.

In limbic epileptic seizures, the oscillation frequency can change both smoothly \cite{Sobayo_etal_IEEEBiomedEng2012} and abruptly \cite{SenhadjiWendling_NeuroClinique2002}. In the presented model, these two ways of frequency evolution are reproduced by smooth changing the resistance of the potentiometer or decreasing/increasing the number of neurons in the network by switching the electronic key. However, the constructed generator may be useful independently of its biological prototype as a source of multi-frequency periodic pulse signals.

The number of neurons examined in this study increased four times compared to \cite{Egorov_CSF2022}. This shows the scaling capabilities of the circuit proposed in \cite{Egorov_CSF2022}. Since the complexity of implementing and considering all possible coupling architectures within the same work would be excessive, we limited ourselves to studying only one coupling architecture (unidirectional ring), which is very important from physiological reasons. 

Compared to earlier studies \cite{Kulminskiy_ND2019,Egorov_ND2022}, in the present experiment all synaptic couplings were analog devices rather than microcontroller based connections using digitized signals; in \cite{Binczak_NN2006} the coupling was analog, but it was linear and therefore not synaptic-like. The synaptic analog coupling, on the one hand, is much more complicated for implementation. But on the other hand, such an implementation is much closer to the biological experiment. The possible next step is switching from hardware realization of FitzHugh--Nagumo model to the Morris--Lecar model which is known as a simplest biologically proven model (some first implementation of which was published in \cite{PatelDeWeerth_1997analogue}).

\section{Acknowledgments}
This work was supported by Russian Science Foundation, grant No.~19-72-10030-P, \verb|https://rscf.ru/project/19-72-10030/|

\bibliographystyle{IEEEtran}
\bibliography{ArxivHippocampalRing}

\end{document}